\documentclass[twocolumn,floatfix,showpacs,showkeys,preprintnumbers,nofootinbib,superscriptaddress]{revtex4}
\usepackage[utf8]{inputenc}
\usepackage[sort&compress]{natbib}
\usepackage{ulem}
\usepackage{bm}
\usepackage{times}
\usepackage{amssymb,amsbsy,amsmath,amsfonts}
\usepackage{graphicx}
\usepackage{float}
\usepackage{color}
\usepackage{morefloats}
\usepackage{rotating}
\usepackage{srcltx}
\usepackage{slashed}
\usepackage{subfigure}
\usepackage{multirow}
\usepackage{verbatim}
\usepackage{hyperref}
\usepackage{tabularx}
\usepackage{adjustbox}

\begin{document}

\title{  $Z_{cs}(3985)$ in next-to-leading-order chiral effective field theory -- the first truncation uncertainty analysis}

\author{Qing-Yu Zhai}
\affiliation{School of Physics, Beihang University, Beijing 102206, China}

\author{Ming-Zhu Liu}
\affiliation{School of Space and Environment, Beihang University, Beijing 102206, China}
\affiliation{School of Physics, Beihang University, Beijing 102206, China}

\author{Jun-Xu Lu}
\email{ljxwohool@buaa.edu.cn}
\affiliation{School of Space and Environment, Beihang University, Beijing 102206, China}
\affiliation{School of Physics, Beihang University, Beijing 102206, China}

\author{Li-Sheng Geng}\email{lisheng.geng@buaa.edu.cn}
\affiliation{School of Physics, Beihang University, Beijing 102206, China}
\affiliation{Beijing Key Laboratory of Advanced Nuclear Materials and Physics, Beihang University, Beijing 102206, China}
\affiliation{School of Physics and Microelectronics, Zhengzhou University, Zhengzhou, Henan 450001, China}

\date{\today}
\begin{abstract}
We revisit the $D_s^-D^{*0}$/$D_s^{*-}D^0$ interaction and the $Z_{cs}(3985)$ state in chiral effective field theory(EFT)  up to the next-to-leading order. We examine the relative importance of the leading-order contact, one-eta exchange,  next-to-leading-order contact, and two-kaon-exchange contributions. We show that the leading-order and next-to-leading-order contact contributions play the most important role such that the $Z_{cs}(3985)$ state qualifies as a $D_s^-D^{*0}$/$D_s^{*-}D^0$ resonance. On the other hand, the weak one-eta-exchange and weakly energy-dependent two-kaon-exchange contributions  are less important in dynamically generating the $Z_{cs}(3985)$ state, indicating that chiral EFT is less predictive in the present situation. Furthermore, we apply the Bayesian method to estimate chiral truncation uncertainties and  find that  they are of similar magnitude as their statistical counterparts. Our study shows  that if $Z_c(3900)$ exists, then SU(3)-flavor symmetry also predicts $Z_{cs}(3985)$ with a certain robustness, i.e., both can be accommodated in the chiral EFT with dominant contact contributions.

\end{abstract}


\maketitle

\section{Introduction}

In 2020 the BESIII Collaboration reported on the existence
of a hidden-charm state with strangeness, $Z_{cs}(3985)$, with a significance of $ 5.3 \sigma$~\cite{BESIII:2020qkh}. The obtained mass and width are:
\begin{eqnarray}
M_{Z_{cs}} &=& 3982.5^{+1.8}_{-2.6}\pm 2.1 ~ {\rm MeV}, \\
\Gamma_{Z_{cs}} &=& 12.8^{+5.3}_{-4.4}\pm 3.0 ~  {\rm MeV}.
\end{eqnarray}
In 2021 the LHCb Collaboration observed two more hidden-charm tetraquark states  with strangeness in the $J/\psi K^+$ invariant mass spectrum of the $B^+\to J/\psi \phi K^+$ decay,  $Z_{cs}(4000)^+$ and $Z_{cs}(4200)^+$~\cite{LHCb:2021uow}. Because the width  of $Z_{cs}(3985)$ and that of $Z_{cs}(4000)$ differ by one order of magnitude and their  production mechanism is also different, they are quite unlikely the same state~\cite{LHCb:2021uow,Meng:2021rdg}. 
As a result, in the present work, we only focus on the $Z_{cs}^-(3985)$ state.

There were quite a number of studies on the likely existence of a $c\bar{c}s\bar{q}$ state in the literature before the experimental discovery~\cite{Ebert:2005nc,Gamermann:2006nm,Lee:2008uy,Liu:2008tn,HidalgoDuque:2012pq,Wu:2018xdi,Voloshin:2019ilw,Ferretti:2020ewe}.
In Ref.~\cite{Wu:2018xdi}, assuming a
compact tetraquark picture and considering the chromomagnetic
interaction, the authors found that the lowest
$c\bar{c}s\bar{q}$ tetraquark is about 100 MeV below the ${D}\bar{D}_{s}$ mass threshold. In the relativistic quark model, the lowest $c\bar{c}s\bar{q}$ state is found to be 10 MeV below the
${D}\bar{D}_{s}$ mass threshold~\cite{Ebert:2005nc}. In
Ref.~\cite{Liu:2008tn} no ${D}^{(\ast)}\bar{D}_{s}^{(\ast)}$ bound
state was found in the one-boson-exchange (OBE) model, where only $\pi$, $\sigma$, $\omega$, and $\rho$ exchanges were considered. In the
chiral unitary approach constrained by the hidden gauge symmetry and
broken SU(4) symmetry~\cite{Gamermann:2006nm}, no resonance or bound
state near the ${D}\bar{D}_{s}$ mass threshold was found. In
Ref.~\cite{HidalgoDuque:2012pq} an effective field theory study with
constraints from heavy quark spin symmetry and  SU(3)-flavor
symmetry predicted several hadronic molecules near the
${D}^{(\ast)}\bar{D}_{s}^{(\ast)}$ thresholds.~\footnote{These results should be
taken with caution because  there $X(3915)$ and 
$Y(4140)$ were assumed to be $D^{\ast}\bar{D}^{\ast}$ and
$\bar{D}_{s}^{\ast}D_{s}^{\ast}$ hadronic molecules with quantum
numbers $J^{PC}=0^{++}$. However, the $J^{PC}$ of $Y(4140)$ has
now been determined to be $1^{++}$~\cite{Tanabashi:2018oca}.} 

After the BESIII discovery~\cite{BESIII:2020qkh}, many new studies were performed and some of the earlier studies were updated. As the $Z_{cs}(3985)$ state lies close to the mass thresholds of
$D_{s}^{-}D^{\ast0}$ and $D_{s}^{\ast-}D^{0}$, the molecular picture has gained a lot of attention~\cite{Yang:2020nrt,Meng:2020ihj,Chen:2020yvq,Sun:2020hjw,Du:2020vwb,Wan:2020oxt,Wang:2020rcx,Wang:2020htx,Xu:2020evn,Guo:2020vmu,Chen:2021erj,Albuquerque:2021tqd,Yan:2021tcp,Chen:2021uou,Ding:2021igr,Baru:2021ddn,Du:2022jjv}. There exist also  competing interpretations. For instance, it has been suggested to   be either a compact tetraquark state~\cite{Wang:2020iqt,Azizi:2020zyq,Wang:2020dgr,Faustov:2021hjs,Albuquerque:2021tqd,Shi:2021jyr,Giron:2021sla,Karliner:2021qok,Yang:2021zhe,Ferretti:2021zis,Maiani:2021tri}, a hadro-charmonium~\cite{Ferretti:2021zis}, a virtual state~\cite{Ortega:2021enc}, a threshold effect~\cite{Ikeno:2020mra},  a reflection effect~\cite{Wang:2020kej}, or a recoupling effect~\cite{Simonov:2020ozp}. A few remarks are in order regarding some of these studies. 
It was shown that the OBE potential does not support the existence of a $D^-_sD^{*0}/D^{*-}D^0$ molecule~\cite{Chen:2020yvq}, while the QCD sum rule approach cannot distinguish between a molecule and a compact tetraquark state~\cite{Wan:2020oxt,Wang:2020rcx,Wang:2020dgr,Albuquerque:2021tqd}. There is no candidate of either molecular nature or compact tetraquark nature for the $Z_{cs}(3985)$ state in the chiral quark model of Ref.~\cite{Jin:2020yjn}.The  existence of the $Z_{cs}(3985)$ state in the OBE model  of Ref.~\cite{Yan:2021tcp} depends on a number of less known factors, such as  the  $\eta$ exchange and
the  scalar meson exchange. In addition to its mass and spin-parity, the decay and production mechanisms~\cite{Cao:2020cfx,Wu:2021ezz,Ikeno:2021mcb,Han:2022fup}, electromagnetic properties~\cite{Ozdem:2021yvo}, and even medium modifications~\cite{Azizi:2020zyq,Sungu:2020zvk} of the $Z_{cs}(3985)$ state have been extensively studied.

In the molecular picture, it is particularly interesting to note that in Ref.~\cite{Wang:2020htx}, using the next-to-leading-order chiral potential, with the two low-energy constants (LECs) and the cutoff determined from the $Z_c(3900)$ data~\cite{Wang:2020dko} and SU(3)-flavor symmetry, it is shown that the $K^+$ recoil-mass spectrum of the $e^+e^-\to K^+(D_s^-D^{*0}+D^{*-}_sD^0)$ reaction can be well described. With the same three parameters, a pole is found at
$(m,\Gamma)=(3982.4^{+4.8}_{-3.4},11.8^{+5.5}_{-5.2})$ MeV, which can be associated with the $Z_{cs}(3985)$ state.  

An effective field theory is a low-energy realization of the underlying (and more fundamental) theory, which satisfies all the relevant global symmetries and allows for model-independent descriptions of physical processes in its validity domain. In the present work, the underlying theory of the strong interaction is quantum chromodynamics (QCD), while the effective field theory is chiral effective field theory (ChEFT). It has been successfully applied to study many low-energy phenomena, in particular, the nucleon-nucleon interaction~\cite{Ordonez:1993tn,Entem:2003ft,Epelbaum:2014sza,Lu:2021gsb}, pioneered by Weinberg~\cite{Weinberg:1990rz,Weinberg:1991um}. For a review of the application of ChEFT in the heavy quark sector, see Ref.~\cite{Meng:2022ozq}. A key  issue in any EFT is that it allows for a controlled estimate of truncation uncertainties in a statistically meaningful way. 

In the past, chiral truncation uncertainties at a given order were usually estimated by the variation of the cutoff needed to regularize the corresponding scattering equation, e.g. in Ref.~\cite{Machleidt:2011zz}. However, as pointed out in Ref.~\cite{Epelbaum:2014efa}, the resulting uncertainties are actually the residue effect of cutoff dependence, which  essentially originate from the missing contact interactions that appear only at even chiral orders. Meanwhile, the uncertainties obtained in this way are somehow arbitrary since they  depend on the employed cutoff range. In Ref.~\cite{Epelbaum:2014efa}, the authors proposed to use the differences between the optimal results obtained at different orders as an estimate of truncation uncertainties, which are referred to as the EKM  uncertainties~\cite{Epelbaum:2014sza}. They are motivated by the expectation that at low energies, truncation uncertainties are dominated by powers of expansion quantities such as momentum or light quark masses. For a short review of the application of the EKM approach in nuclear physics, see Ref.~\cite{Epelbaum:2019kcf}. The EKM uncertainties are by construction smaller order by order (given that the EFT converges). However, since only the known  contributions are taken into account, a statistical interpretation of this algorithm is missing.  More recently, a new method based on the Bayesian model has been proposed~\cite{Furnstahl:2015rha,Melendez:2017phj,Melendez:2019izc}, in which by encoding the expectation of the chiral expansion coefficients in a ``prior probability density function'', the truncation uncertainties for a certain degree-of-belief (DoB) can be  determined by integrating out the coefficients of omitted orders. The new Bayesian approach and its modified version have been applied in the latest studies of pion-nucleon scattering~\cite{Reinert:2020mcu}, NN and 3N scattering~\cite{Epelbaum:2019zqc,Lu:2021gsb,Volkotrub:2020lsr} and other nuclear physics observables ~\cite{Zhang:2015ajn,Zhang:2019odg,Maris:2020qne,Djarv:2021pjc}. 

In this work, for the first time, we apply the Bayesian model to the heavy quark sector and to studies of exotic hadrons. In particular, we study the $D_s^-D^{*0}$/$D_s^{*-}D^0$ scattering and the related $Z_{cs}(3985)$ state in chiral effective field theory (ChEFT) up to the next-to-leading order (NLO). 
 In particular, we focus on the decomposition of the NLO chiral potential and apply the Bayesian model to study the chiral truncation uncertainties. We show that the expansion converges well up to the breakdown scale of the cutoff and even taking into account chiral truncation uncertainties, the
state remains as a resonance, supporting its identification as the $Z_{cs}(3985)$ state~\cite{Wang:2020htx}.

This work is organized as follows. We briefly explain the ChEFT approach and the Bayesian model in Sec.~II. Results and discussions are given in Sec.III, followed by a short summary in the last section.

\section{Theoretical Formalism}
\label{sec:obe}

In this section, we introduce the chiral potentials up to the next-to-leading order and explain the Bayesian model adopted in the present work.

\subsection{Chiral potentials up to  next-to-leading order}

\begin{figure}[hptb]
\centering
\includegraphics[width=0.3\textwidth]{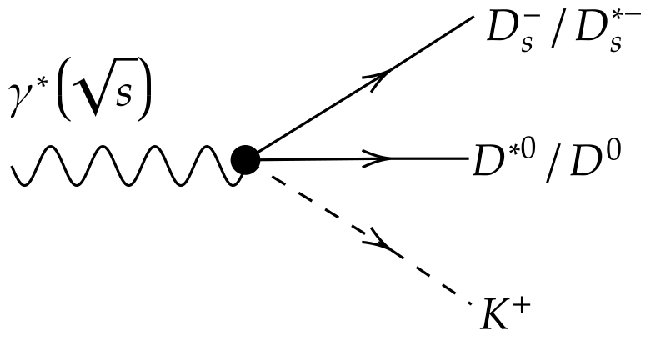}\hspace{-5mm}\\
\includegraphics[width=0.3\textwidth]{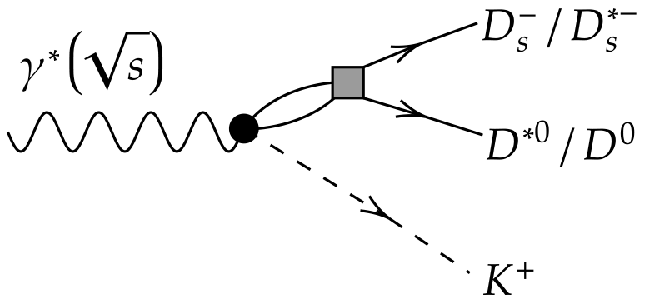}\\ \vspace{0mm}
\centering
\caption{Feynman diagram for the $e^+e^-\to K^+(D_s^-D^{*0}+D_s^{*-}D^0)$ reaction: tree-level (upper panel) and rescattering process (lower panel). }\label{fig:feynman}
\end{figure}
In this work, we follow Refs.~\cite{Wang:2020htx,Wang:2020dko} and describe the $e^+e^-\rightarrow K^+(D_s^-D^{*0}-D_s^{*-}D^0)$ reaction in two steps as shown in Fig.~\ref{fig:feynman}: the decay of a virtual photon into $K^+(D_s^-D^{*0}+D_s^{*-}D^0)$~\footnote{It should be noted that the virtual photon decaying into   $K^+(D_s^-D^{*0}+D_s^{*-}D^0)$ can also proceed via the triangle mechanism (see, e.g., Refs.~\cite{Yang:2020nrt,Du:2022jjv}. However, the currently limited experimental data are not enough to unambiguously distinguish these two mechanisms. As a result, following Refs.~\cite{Wang:2020htx,Wang:2020dko}, we only consider the mechanism shown in Fig.~1.   } and the rescattering of $D_s^-D^{*0}/D_s^{*-}D^0$. The amplitude of the whole process $\mathcal{U}(E,\pmb{p})$ can be obtained by solving the following Lippmann-Schwinger equation

\begin{align}
  \mathcal{U}(E,\pmb{p})=\mathcal{M}(E,\pmb{p})+\int\frac{\mathrm{d}^{3}\pmb{q}}{(2\pi)^{3}}\mathcal{V}(E,\pmb{p},\pmb{q})\frac{2\mu}{\pmb{p}^{2}-\pmb{q}^{2}+i\epsilon}\mathcal{U}(E,\pmb{q}),
\end{align}
where $E$, $\pmb{p}$ and $\mu$ are the energy, momentum and reduced mass of the $D^-_s D^{*0}$/$D^{*-}_sD^0$ system in Fig.~\ref{fig:feynman} respectively and $|\pmb{p}|=\sqrt{2\mu(E-m_{th})}$ with $m_{th}$ the threshold of the system.
  $\mathcal{M}(E,\pmb{p})$ is the photon decay amplitude, which can be described by the following effective Lagrangian
\begin{align}
    \mathcal{L}_{\gamma^{*}\varphi VP}&=g_{\gamma}\mathcal{F}^{\mu\nu}[(\tilde{P}^{\ast\dagger}_{\mu}u_{\nu}P^{\dagger}\notag\\&-\tilde{P}^{\ast\dagger}_{\nu}u_{\mu}P^{\dagger})-(\tilde{P}^{\dagger}u_{\mu}P_{\nu}^{\ast\dagger}-\tilde{P}^{\dagger}u_{\nu}P_{\mu}^{\ast\dagger})]+\mathrm{H.c.},
\end{align}
where $g_{\gamma}$ denotes the effective coupling constant, $\mathcal{F}^{\mu\nu}$ is the field strength tensor of the virtual photon,  $P^{(\ast)}=(D^{0(\ast)},D^{+(\ast)},D^{+(\ast)}_s)$ collect the charmed vector/pseudoscalar meson fields with $\tilde{P}^{(\ast)}$ the corresponding anti-meson fields. The $u_{\nu}=\frac{i}{2}\{\xi^{\dagger}, \partial_{\mu}\xi\}$ is the axial-vector field in which $\xi=$exp($\frac{i\phi}{2f_{\phi}}$) with $\phi$ the pseudoscalar octet and $f_{\phi}$ the decay constant. 

The potential $\mathcal{V}(p,q)$ consists of four parts: the LO and NLO contact potential, the one-eta-exchange (OEE) potential, and the two-kaon-exchange (TKE) potential. Their explicit expressions~\cite{Wang:2020htx}  are 
\begin{align}
    &\mathcal{V}_{\mathrm{ct}}=\tilde{C}_{\mathrm{S}}+C_{\mathrm{S}}(p^{2}+p'^{2}),\\
    &\mathcal{V}_{\mathrm{OEE}}=-\frac{g^{2}}{18f_{\eta}^{2}} \int \mathrm{d}\phi \mathrm{d}(\mathrm{cos}(\theta)) \frac{\pmb{q}^{2}}{\pmb{q}^{2}+m_{\eta}^{2}},\\
    &\mathcal{V}_{\mathrm{TKE}}=\int \mathrm{d}\phi \mathrm{d}(\mathrm{cos}(\theta)) V_{1},
\end{align}
where $V_{1}$ is
\begin{align}
    V_{1}=&-\frac{24(4g^{2}+1)m_{K}^{2}+(38g^{2}+5)\pmb{q}^{2}}{2304\pi^{2}f_{K}^{4}}\notag
    \\&+\frac{6(6g^{2}+1)m_{K}^{2}+(10g^{2}+1)\pmb{q}^{2}}{768\pi^{2}f_{K}^{4}}\ln\frac{m_{K}^{2}}{(4\pi f_{K})^{2}}\notag
    \\&+\frac{4(4g^{2}+1)m_{K}^{2}+(10g^{2}+1)\pmb{q}^{2}}{384\pi^{2}f_{K}^{4}y}\varpi \mathrm{arctan}\frac{y}{\varpi}.
\end{align}
As was demonstrated in Ref.~\cite{Wang:2020dko}, the S-D mixing effect is insignificant and in the present work, we neglect the D-wave interaction as was done in Ref.~\cite{Wang:2020htx}. In the above chiral potential, $f_\eta=116$ MeV, $f_K=113$ MeV,  $g=0.57$, $p$ and $p'$ represent the momenta of initial and final state in the center-of-mass (c.m) system, $\pmb{q}=\pmb{p}'-\pmb{p}$, $m_\eta$ and $m_K$ are the masses of eta and kaon, $\theta$ is the scattering angle in the c.m system of $D_s^-D^{*0}/D_s^{*-}D^0$, $y=\sqrt{2pp'cos\theta-p^{2}-p'^{2}}$ and $\varpi=\sqrt{\pmb{q}^{2}+4m_{K}^{2}}$. A  Gaussian form factor,  $\mathrm{exp}(-p'^{2}/\Lambda^{2}-p^{2}/\Lambda^{2})$, is multiplied to the chiral potential to remove the ultraviolet divergence in solving the Lippmann-Schwinger equation.

It should be noted that the OEE and TKE potentials are pure predictions, while the two LECs $\tilde{C}_S$ and $C_S$ need to be determined by fitting to experimental data. If the OEE and TKE potentials are attractive and strong enough to allow for formations of bound states or resonances, then the ChEFT is predictive. Otherwise, it is not. It is one of our purposes to check whether this is the case for the $D_s^-D^{*0}$/$D_s^{*-}D^0$ system.   

\subsection{Bayesian model}
In this subsection, we briefly explain how the Bayesian model can be extended to study the likely existence of resonances. For details, please refer to Refs.~\cite{Furnstahl:2015rha,Melendez:2017phj,Melendez:2019izc}.

Consider the EFT expansion for an observable $X$ at fixed kinematics,
\begin{align}
X=X_{\mathrm{ref}}\sum_{n=0}^{\infty}c_{n}Q^{n}=X^{(0)}+\Delta X^{(2)}+\dots,
\end{align}
where $\{c_{n}\}$ are dimensionless expansion coefficients. $X_{\mathrm{ref}}$ is a natural-sized  $X$ taken as a reference which is suggested~\cite{Epelbaum:2019zqc} to be
\begin{align}
    X_{\rm{ref}}=\text{Max}\{|X^{\rm{LO}}|,\frac{|X^{\mathrm{LO}}-X^{\mathrm{NLO}}|}{Q^2} \}.
\end{align}
$Q$ is the expansion parameter, which is assumed to take the form~\cite{Epelbaum:2014efa,Epelbaum:2014sza,Melendez:2017phj,Furnstahl:2015rha}
\begin{align}
    Q=\text{Max}\{ \frac{p}{\Lambda}, \frac{m_{K/\eta}}{\Lambda} \},
\end{align}
with $p$ the momentum in the c.m. frame, and $\Lambda$ the breakdown scale. However, in the present work, as we show later, the optimal cutoff obtained is less than 0.4 GeV. Therefore, we could not use the kaon mass or eta mass to estimate the expansion parameter. Instead we use an effective mass of 200 MeV to put a lower limit on the expansion parameter, which is about 0.5.
$\Delta X^{(2)}$ is the difference between the NLO result and the LO result.
If the series is truncated at order $k$, then the truncation uncertainty is defined as the sum of the contributions of chiral orders higher than $k$
\begin{align}
    \Delta_{k}=\sum_{n=k+1}^{\infty}c_{n}Q^{n},
\end{align}
removing the dimensional $X_{\mathrm{ref}}$. To estimate  $\Delta_{k}$, one has to start from the known $\{c_{n}\}(n \leq k)$. That is, one needs to specify the probability distribution function $p(\Delta_{k}|c_{0},c_{1},\dots,c_{k})$, with which  the degree-of-belief (DoB) intervals can be calculated  as
\begin{align}
    p\%=\int_{-d_{k}^{(p)}}^{d_{k}^{(p)}}p(\Delta_{k}|c_{0},c_{1},\dots,c_{k})\mathrm{d}\Delta_{k},
\end{align}
with $d_k^{(p)}$ the integral interval corresponding to the DoB $p\%$.
Usually one chooses $p\%=68\%$. The truncation uncertainty for the observable X is then $\Delta X^{(k)}=X_{\text{ref}}d_k^{(p)}$, which means that the possibility of the value of observable $X$ in $(X^{(k)}-\Delta X^{(k)},X^{(k)}+\Delta X^{(k)})$ is $p\%$. 

Following Ref.~\cite{Melendez:2017phj}, we ultilize a Gaussian prior probability distribution function (pdf) with a common hyperparameter $\bar{c}$ for the expansion coefficients $c_i$ as
\begin{equation}
    \text{p}(c_i|\bar{c})=\frac{1}{\sqrt{2\pi}\bar{c}}~e^{-c_i^2/2\bar{c}^2}.
\end{equation}
where $\bar{c}$ itself follows the log-uniform pdf~\cite{Schindler:2008fh} as
\begin{equation}
    \text{pr}(\bar{c})=\frac{1}{\ln(\bar{c}_{>}/\bar{c}_{<})}\frac{1}{\bar{c}}\theta(\bar{c}-\bar{c}_{<})\theta(\bar{c}_{>}-\bar{c}),
\end{equation}
where $\bar{c}_{<}$($\bar{c}_{>}$) is the lower(upper) limit of $\bar{c}$ and $\theta(x)$ here denotes the step function. 
The posterior pdf $p(\Delta_{k}|c_{0},c_{1},\dots,c_{k})$ then takes the following form~\cite{Epelbaum:2019zqc}
\begin{align}
    p(\Delta_{k}|&c_{0},c_{1},\dots,c_{k})=\frac{1}{\sqrt{\pi \bar{q}^{2} \pmb{c}_{k}^{2}}}\Bigg(\frac{\pmb{c}_{k}^{2}}{\pmb{c}_{k}^{2}+\Delta^{2}/\bar{q}^{2}}\Bigg)^{k/2} \notag \\
    &\quad \frac{\Gamma \big[\frac{k}{2},\frac{1}{2\bar{c}_{>}^{2}}\big( \pmb{c}_{k}^{2}+\frac{\Delta^{2}}{\bar{q}^{2}} \big)\big] -
    \Gamma \big[\frac{k}{2},\frac{1}{2\bar{c}_{<}^{2}}\big( \pmb{c}_{k}^{2}+\frac{\Delta^{2}}{\bar{q}^{2}} \big)\big]}{\Gamma \big[\frac{k-1}{2},\frac{\pmb{c}_{k}^{2}}{2\bar{c}_{>}^{2}}\big]-\Gamma \big[\frac{k-1}{2},\frac{\pmb{c}_{k}^{2}}{2\bar{c}_{<}^{2}}\big]},
\end{align}
where $\bar{q}^{2} \equiv \sum_{i=k+1}^{k+h}Q^{2i}$ with $h$ the next highest order to be taken into account, $\pmb{c}_{k}^{2} \equiv \sum_{i \in A}c_{i}^{2}$, $A \equiv \{n\in \mathbb{N}_{0}|n\leq k \wedge n\neq 1 \wedge n\neq m \}$ with $c_m$ the expansion coefficient corresponding to the overall scale $X_{\text{ref}}$ defined above. In the present work, we take $k=2$, $h=10$, $\bar{c}_{<}=0.5$ and $\bar{c}_{>}=10.0$ following Ref.~\cite{Melendez:2017phj}.

\section{Numerical results and Discussions}
\label{sec:pre}
In this section, we study the decomposition of the chiral potential up to NLO and perform two types of truncation uncertainty analyses.  In Table \ref{tab:mass}, we tabulate the masses of the relevant particles.

\begin{table}[htpb]
    \centering
        \caption{Masses of particles~\cite{Tanabashi:2018oca} relevant to the present study (in units of GeV). }\label{tab:mass}
    \begin{tabular}{ c c c c }
    \hline\hline
    particle & mass & particle & mass \\
    \hline
    $D_s^{*\pm}$ & 2.1122 & $D_{s}^{\pm}$ & 1.9683\\
    $D^0$ & 1.86484 & $D^{\pm}$ & 1.86965 \\
     $D^{*0}$ & 2.00685 & ${D}^{*\pm}$  &2.01026  \\
     $K^{\pm}$  &0.493677 & $\eta$ & 0.54786   \\
     $K^{0}$  &0.497611 &  &    \\
    \hline\hline
    \end{tabular}
\end{table}

\subsection{Decomposition of the chiral potential}
First, we examine the relative importance of the different terms of the chiral potential up to  NLO. In the nucleon-nucleon system, it is well known that the one-pion exchange(OPE) potential provides the longest-range nuclear force, while the two-pion exchange(TPE) contributions describe the intermediate part. As a result, for higher partial waves, e.g., those with angular momentum larger than 3 or 4, the OPE plus TPE potentials can already describe well the corresponding partial wave phaseshifts~\cite{Kaiser:1997mw,Xiao:2020ozd}.  On the other hand, for those channels of low angular momenta, the short-range contributions encoded in the  low-energy constants (LECs) are important, to such an extent that for low energy regions the pions can be integrated out~\cite{vanKolck:1997ut}. The above discussion is relevant because if the OPE and TPE contributions are dominant, the theory is more predictive. Otherwise, one has to rely on either experimental data or lattice QCD data or phenomenology (e.g., resonance saturation) to determine the relevant LECs. This is particularly relevant to the $D_s^-D^{*0}/D_s^{*-}D^0$ interaction and the corresponding $Z_{cs}(3985)$ state. This is because if the LO OEE potential or the NLO TKE potential is attractive enough, then without relying on other inputs, the ChEFT approach is predictive by  itself. Otherwise, certain experimental inputs are needed to predict the existence of $Z_{cs}(3985)$.

\begin{figure}[htpb]
\centering
\includegraphics[width=\linewidth]{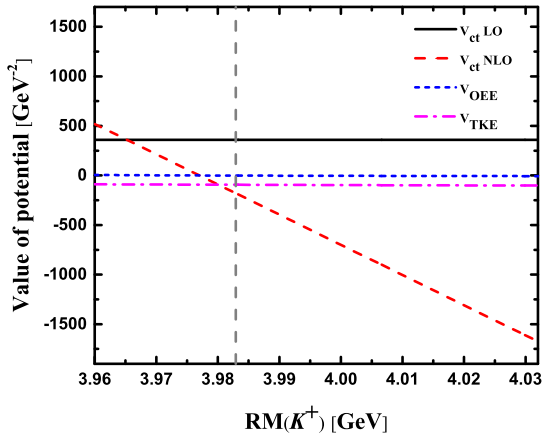}
\centering
\caption{Decomposition of the chiral potential up to NLO as a function of $K^+$ recoil-mass RM($K^+$). The vertical dashed line denotes the real part of the pole position predicted in Ref.~\cite{Wang:2020htx}.}\label{fig:decom}
\end{figure}

In Fig.~\ref{fig:decom}
with the two LECs and the cutoff determined in Ref.~\cite{Wang:2020dko} by fitting to the $Z_c(3900)$ data, i.e., $\tilde{C}_S=3.6\times10^{2}\,\mathrm{GeV}^{-2}$, $C_S=-76.9\times10^{2}\,\mathrm{GeV}^{-4}$, and $\Lambda=0.33\,\mathrm{GeV}$, we compare the LO contact, LO OEE, NLO contact, and NLO TKE contributions around the $D_s^-D^{*0}$/$D_s^{*-}D^0$ threshold. It is clear that the OEE contribution is negligible compared to the LO contact term. In addition, in the energy region studied, the TKE contribution is nearly energy independent and is only about one quarter of  the LO contact contribution. From this, we conclude that the OEE or TKE contribution alone is not  enough to dynamically generate the $Z_{cs}(3985)$ state as a $D_s^-D^{*0}$/$D_s^{*-}D^0$ resonance. On the other hand, the NLO contact contribution is zero at threshold but increases quickly as one moves away from the threshold. At the pole position, its size is about half  that of the LO contact contribution.  50 MeV above the threshold, it  already becomes four times stronger. As a matter of fact, the NLO contributions (both contact and TKE) are responsible for the appearance of a resonant $D_s^-D^{*0}-D_s^{*-}D^0$ state,  because with  the energy-independent LO contribution, only bound or virtual states can emerge.
 
It is interesting to note that the LO contact potential is repulsive while the OEE potential is attractive but negligible in magnitude. The TKE potential is moderately attractive while the NLO contact potential is attractive above the threshold but repulsive below the threshold.

\subsection{Truncation uncertainty analysis}
We perform two types of truncation uncertainty analyses. One is based on the chiral potential, which itself is not a physical observable.  The other is based on the $K^+$ recoil-mass spectrum, which is a more proper observable. 

\subsubsection{Uncertainty analysis based on chiral potential}
\begin{figure}[htpb]
\centering
\includegraphics[width=\linewidth]{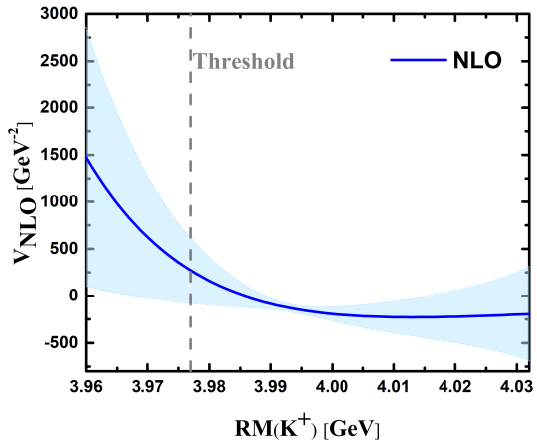}
\centering
\caption{NLO chiral potential~\cite{Wang:2020htx}  as a function of the $\bar{D}_sD^*$ invariant mass. The band represent chiral truncation uncertainties obtained by the Bayesian model for a DoB of 68\%.}
\label{fig:verror}
\end{figure}

In Fig.~\ref{fig:verror}, we plot the NLO chiral potential of Ref.~\cite{Wang:2020htx} as a function of the $D_s^-D^{*0}/D_s^{*-}D^0$ invariant mass and the corresponding truncation uncertainty for a DoB of 68\%. The chiral truncation uncertainty is relatively small close to the threshold, but becomes larger around the breakdown scale. We need to mention that for the potential below the threshold, we have replaced $p$ with $|p|$ in the calculation of the expansion parameter $Q$. One can translate the truncation uncertainties of the chiral potential shown in Fig.~\ref{fig:verror} to those of the two LECs, $\tilde{C}_S$ and $C_S$. They now become $\tilde{C}_S=3.60^{+1.2+0.5}_{-1.2-0.5}$ GeV$^{-2}$ and $C_S=-76.9^{+6.2+10.0}_{-6.2-10.0}$ GeV$^{-4}$, where the first uncertainty is statistical and the second is systematic originating from chiral truncations. 

\begin{figure}[htpb]
\centering
\includegraphics[width=\linewidth]{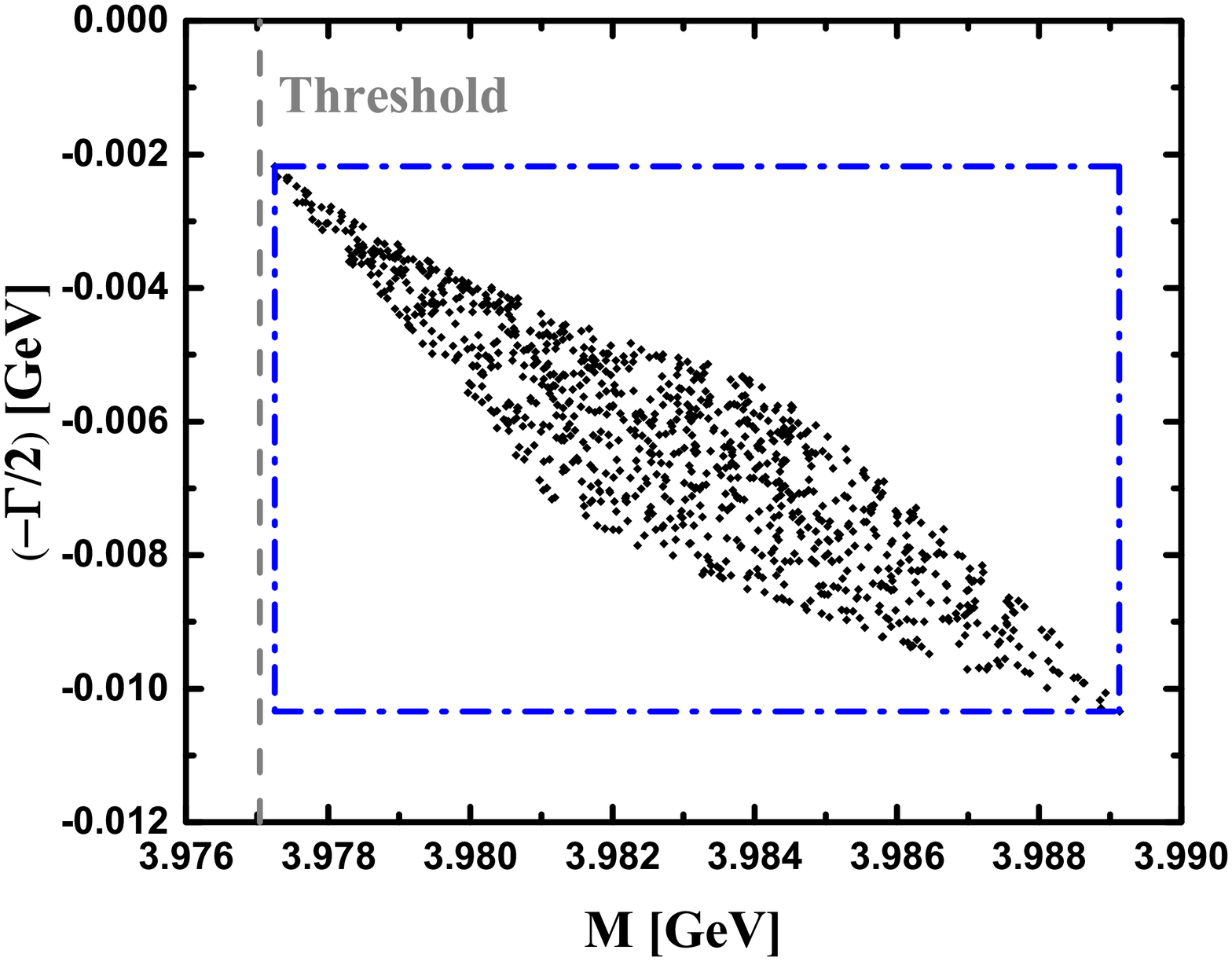}
\centering
\caption{Pole positions corresponding to $Z_{cs}(3985)$ with systematic and statistical uncertainties for the NLO chiral potential of Ref.~\cite{Wang:2020htx}. The blue dash-dotted frame defines the uncertainty boundaries.}
\label{fig:pole}
\end{figure}

With the NLO chiral potential and the corresponding truncation uncertainty as well as statistical uncertainty, 
we search for poles on the second Riemann sheet. The resulting positions are shown in  Fig.~\ref{fig:pole}.~\footnote{The pole positions are calculated using the LECs  randomly sampled within the one $\sigma$ regions around their central values.} Clearly, even with the truncation uncertainties taken into account,  ChEFT still supports the interpretation of the $Z_{cs}(3985)$ state as a $D_s^-D^{*0}$/$D_s^{*-}D^0$ resonance. More specifically, with both uncertainties, we obtain the position as $(m,\Gamma)=(3982.4_{-5.2}^{+6.7}, 11.8_{-7.4}^{+8.9})$ MeV, which should be compared with that obtained in Ref.~\cite{Wang:2020htx}, $(m,\Gamma)=(3982.4^{+4.8}_{-3.4},11.8^{+5.5}_{-5.2})$ MeV. Clearly even compared to the relatively large statistical uncertainty, the systematic uncertainty is sizable. 

\subsubsection{Uncertainty analysis based on  $K^+$-recoil mass spectrum}
\begin{figure}[htpb]
\centering
\includegraphics[width=\linewidth]{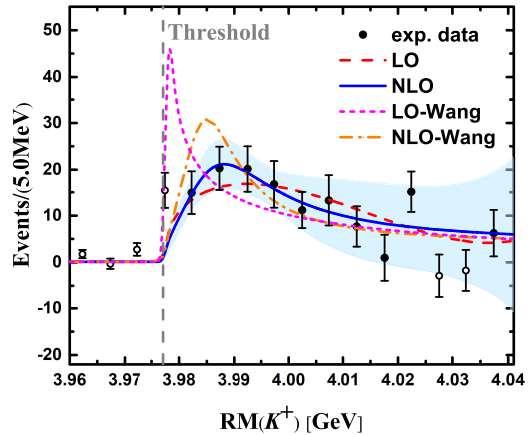}
\centering
\caption{$K^+$ recoil-mass spectrum of the $e^+e^-\to K^+(D_s^-D^{*0}+D_s^{*-}D^0)$ reaction. The light blue band represents chiral truncation uncertainties obtained using the Bayesian model for a DoB of 68\%. The curves labeled by ``LO-Wang'' and ``NLO-Wang" are the LO and NLO results of  Ref.~\cite{Wang:2020htx}. The solid dots associated with errors are experimental data after removing the combinatorial backgrounds from Ref.~\cite{BESIII:2020qkh} }
\label{fig:newfit}
\end{figure}

The leading order potential of Ref.~\cite{Wang:2020htx} describes the $K^+$-recoil spectrum very badly (see Fig.~\ref{fig:newfit}), therefore 
to perform a proper truncation uncertainty analysis, we perform a new fit to the experimental data of Ref.~\cite{BESIII:2020qkh}. At LO, we have two LECs to determine, i.e., $\tilde{C}_S$ and $\Lambda$. At NLO, the LEC $C_S$ also contributes. By fitting to the experimental data shown in Fig.~\ref{fig:newfit}~\footnote{As a matter of fact, only the data points denoted by solid points are fitted, while those denoted by open circles are not. This choice is motivated by the following observation. Among the three data points neglected, two of them have negative central values and the one at threshold cannot be simultaneously described together with the others.}, we obtain the LECs given in Table~\ref{tab:fits} together with the corresponding $\chi^2/\mathrm{d.o.f.}$. The light blue band is generated using the Bayesian method explained in Sec.IIB with $\Lambda=0.36$. It should be mentioned that in the spirit of the Bayesian model, we have fixed the cutoff at its central value in obtaining the statistical uncertainties of the LECs. For the sake of comparison, we also show the LO and NLO predictions of Ref.~\cite{Wang:2020htx}. It should be noted that the experimental data we fitted are taken from Fig.~4 of Ref.~\cite{BESIII:2020qkh} where the wrong-sign combinations of $D_s^-$ and $K^-$ candidates and the excited $D_{s}^{**+}$ contributions have been subtracted.  Our fits indicate that there is no need to consider $K$ rescattering from either $D^{(*)-}_{s}$
or $D^{(*)0}$, in agreement with Refs.~\cite{Du:2022jjv,Baru:2021ddn,Wang:2020htx,Du:2020vwb,Yang:2020nrt}. Once more precise data become available, one may need to examine this part of final-state interactions in more detail.

\begin{table*}[htpb]
    \centering
        \caption{LECs of the LO and NLO chiral potential and the cutoff needed to regularize the chiral potential obtained by fitting to the BESIII data shown in Fig.5 (the solid points). $\tilde{C}_S$ is in units of $10^2$ GeV$^{-2}$, $C_S$ is in units of $10^2$ GeV$^{-4}$, and $\Lambda$ is in units of GeV. For the NLO LECs, the first uncertainty is statistical (originating from the uncertainties of the BESIII data) and the second is systematic (originating from chiral truncation uncertainties, see the main text for details).}\label{tab:fits}
    \setlength{\tabcolsep}{4.5mm}{
    \begin{tabular}{ c c c c c }
    \hline\hline
     & $\tilde{C}_S$ & $C_S$ & $\Lambda$ & $\chi^2/\mathrm{d.o.f.}$ \\
    \hline
    LO & $13.38_{-0.28}^{+0.28}$ & $\cdots$ & $0.48_{-0.01}^{+0.01}$ & $0.965$ \\
    NLO & $2.33_{-1.45-0.93}^{+1.45+0.93}$ & $-46.80_{-11.56-7.04}^{+11.56+7.04}$ & $0.36_{-0.14}^{+0.14}$ & $0.662$ \\
    \hline\hline
    \end{tabular}}
\end{table*}

A few things are noteworthy. First, the NLO description of the experimental data is quantitatively better than its LO counterpart. Nonetheless, both the LO and NLO fits can be viewed as reasonable  given the relatively large experimental uncertainties. We note that in the literature, EFT studies have been performed at both LO and NLO, and reasonable descriptions of the experimental data have been claimed. Second, the truncation uncertainties increase quickly as one moves away from the threshold. This is mainly because of the relatively small optimal cutoff of 0.36 GeV~\footnote{This value or the value of 0.33 GeV obtained in Ref.~\cite{Wang:2020dko} is a bit small if one considers that the eta and kaon masses are about 0.5 GeV. On the other hand, considering that the OEE and TKE contributions are relatively small and almost energy independent  and therefore can be absorbed into the LO contact contribution, the relatively small cutoff is acceptable.} obtained at NLO, which implies that the NLO ChEFT can be trusted at most up to $\sqrt{s}_{D_s^-D^{*0}/D_s^{*-}D^0}\approx4.041$ GeV. With the parameters given in Table \ref{tab:fits}, we search for poles on the complex plane. There is no bound state or virtual state at LO, but we find a resonant state at NLO. The position is $(M,\Gamma)=(3983.5_{-6.5}^{+9.8}, 21.3_{-12.5}^{+18.5})$ MeV, where the uncertainties include both systematic and statistical ones. We note that the pole position is consistent with that of Ref.~\cite{Wang:2020htx} within the relatively large uncertainty.

\section{Summary and outlook}
\label{sec:conclusions}
We revisited the $e^+e^-\to K^+(D_s^-D^{*0}+D_s^{*-}D^0)$ reaction in chiral effective field theory up to the next-to-leading order in the single-channel approximation. We first examined the relative importance of various contributions  in the chiral expansion and showed explicitly  the important role played by the leading order and next-to-leading-order contact contributions  in describing the BESIII data. This demonstrates that for the particular case studied, the ChEFT approach is not very predictive by itself. Experimental inputs or other symmetries  are needed to predict the existence of the $Z_{cs}(3985)$ state. This was done in Ref.~\cite{Wang:2020htx} using the $Z_c(3900)$ data and SU(3) symmetry. Then we applied the Bayesian method to estimate chiral truncation uncertainties. We performed  two types of uncertainty analyses, based on either the chiral potential or the event distribution. For the latter, we have refitted the BESIII data. Our results showed that because of the relatively small cutoff obtained from the fitting, the chiral EFT approach can only be trusted up to 50 MeV above the $D_s^-D^{*0}$/$D_s^{*-}D^0$ threshold. In either case, our results showed that even with chiral truncation uncertainties taken into account, the $Z_{cs}(3985)$ state remains a robust $D_s^-D^{*0}$/$D_s^{*-}D^0$ resonance. 

To the best of our knowledge, the present study constitutes the first exploration of the Bayesian model in studies of exotic hadrons. It is shown that a proper and statistically meaningful uncertainty analysis can be performed. Such an analysis gives us more confidence in the prediction and/or results of the chiral effective field theory studies. We hope that the present study can stimulate more similar studies in the heavy flavor sector in the future.  

\section{Acknowledgments}
This work is partly supported by the National
Natural Science Foundation of China under Grants Nos.11735003,
11975041, 11961141004 and the fundamental Research
Funds for the Central Universities. Ming-Zhu Liu acknowledges support from the National Natural Science Foundation of China under Grant No.12105007.  Junxu Lu acknowledges support from the National Natural Science Foundation of China under Grant No.12105006 and China Postdoctoral Science Foundation under Grant No. 2021M690008.

\newpage

\bibliography{mybib}

\end{document}